\documentclass{article}
\usepackage[utf8]{inputenc}

\usepackage{fullpage}
\usepackage{amsmath}
\usepackage{amssymb}
\usepackage{lipsum}
\usepackage{amssymb,amsthm,amsmath}
\usepackage{amsfonts}
\usepackage{esint}
\usepackage{amsfonts}
\usepackage{amsthm}
\usepackage{graphicx}
\usepackage{todonotes}
\usepackage{xcolor}
\usepackage{amsmath,blkarray}
\usepackage[hidelinks]{hyperref}
\usepackage{caption}
\usepackage[labelfont=bf]{caption}
\usepackage[capitalise]{cleveref}
\usepackage{comment}
\usepackage{tikz}
\usetikzlibrary{automata,arrows,positioning,calc}

\def\dt{\Delta t}

\title{An Agent-based Model for Competitive Agents}
\author{Mohammad Daneshvar\thanks{Dept. of Mathematics, Houston Community College, Houston, TX, 77002, USA,
(mohammad.daneshvar@hccs.edu)}\quad 
Mandana Delavari\thanks{Dept. of Mathematics, University of Houston, Houston, TX 77204, mdelavari@uh.edu}}

\begin{document}
\date{}
\maketitle

\begin{abstract}
In this paper, we analyze the behavior of a multi-agent system driven by the interactions of agents within a competitive environment. To achieve this, we describe the transition probabilities that underlie the system's stochastic nature. We also derive the Fokker-Planck equations for the density distribution of the number of agents in the system and solve these equations for specific cases.
\end{abstract}


\section{Introduction}

Continuous-time Markov chains have been employed for decades to model a broad spectrum of stochastic systems, including queuing systems (e.g., \cite{danesh}) and financial markets (e.g., \cite{wang, tesfatsion2006agent}).
These models often represent agent behavior in interactive environments, where local and global interaction rules are used to simulate various physical processes (e.g., see \cite{1,2,3} for examples). A key question in the analysis of these models is how to derive the transient or stationary probability distributions that capture the system’s evolving dynamics or long-term behavior.\\

In this paper, we develope a straightforward stochastic agent-based model for the analysis of agents displaying competitive behavior, striving to survive within a competitive environment. This model has applications across applied finance and social science (see \cite{4}). For instance, in financial markets, firms compete to attract more customers and clients; job market participants frequently switch employers to better fulfill their financial needs; governments work to strengthen their economies, and so forth.\\

In the subsequent section, we begin with a microscopic model where numerous groups or agents exist, each containing a finite number of subagents. At each time increment, each subagent associates with a single group. Subagents migrate between groups at specified rates. More formally, we consider $n$ competing agents indexed as $(X_i(t))_{i=1}^{n}$, where $X_i(t)$ represents the wealth of agent $i$ at time $t$. In the discrete-time model, the competition between agents $i$ and $j$ is depicted as follows: at time step $t+1$, agent $i$ consumes one unit of the wealth of agent $j$ with a specified probability denoted as $\kappa_{ij}$. This probability often depends on the wealth of agents $i$ and $j$ at time step $t$. If $X_i(t)=0$, it implies that agent $i$ does not engage in any competition at step $t+1$, meaning $\kappa_{ij}=\kappa_{ji}=0$ for all $j$. Essentially, if an agent's wealth depletes entirely, it remains in this state and effectively exits the model. Additionally, the assumption is made that agents do not compete against themselves, i.e., $\kappa_{ii}=0$ for all $i$. Furthermore, the model assumes conservation, meaning no wealth flows in or out, thus maintaining a constant total wealth $\sum\limits_{i=1}^{n} X_i(t)$ over time.\\

Through a scaling approach, a mesoscopic model is derived in section \ref{sec3}, as the number of agents and agent types increases. We then obtain Fokker-Planck equations describing the transient probability distribution of the stochastic system in the continuum model and provide solutions for distinct cases involving linear consumption rates.

\section{A Microscopic Model}
\label{sec2}
Let $\vec{X}(t)=\left(X_1(t),\dots,X_n(t)\right)$ denote the wealth vector, and let $\vec{x}=(x_1, \dots, x_n) \in (\mathbb{Z^*})^n$ represent a specific state. We denote 
\begin{equation}
\label{eq:1}
\text{Pr} \left(\vec{X}(t+1) =(x_1, \ldots , x_{i}+1,\dots, x_{j}-1, \ldots, x_n) |\vec{X}(t)=\vec{x} \right),
\end{equation}
as the probability of agent $i$ consuming one unit of agent $j$'s wealth at time step $t+1$, given the system is in state $x$ at time step $t$. We introduce the jump vector $e_{ij}$, where its $i$-th element is $1$, its $j$-th element is $-1$, and the remaining elements are zero. It's worth noting that $e_{ij}=-e_{ji}$. This model is perceived as a discrete-time discrete-state time-homogeneous Markov chain, and its stochastic dynamics are determined by the following probabilities:


\begin{equation}
\label{eq:2}
\text{Pr}\left(\vec{X}(t+1) = \vec{x}+e_{ij}|\vec{X}(t)=\vec{x} \right) = \nu_{i,j}(\vec{x})\quad \text{for} \,\; i, j=1,\dots,n, \quad \text{and} \;\, i\neq j.
\end{equation}
Here, $\nu_{ij}$ denotes the probability that, at step $t+1$, agents $i$ and $j$ are chosen and agent $i$ subsequently consumes agent $j$. Thus,
\begin{equation}
\label{eq:3}
\nu_{ij} = \frac{1}{{n\choose2}} \times \kappa_{ij}.
\end{equation}

Equation \eqref{eq:2} describes the flow of wealth between the agents.\\

%
The consumption rate, $\nu_{ij}$, depends on the wealth of agents $i$ and $j$. We assume $\nu_{ij}(\vec{x})=0$ for all $j$ whenever $x_i=0$. This corresponds to the situation in that agent $i$ has lost its total wealth and is removed from the competition. \\

The stochastic model defined constitutes a discrete-time, discrete-state Markov chain $\vec{X}(t)$ with values on
$(\mathbb{Z}^*)^n$. The dynamics of the model are given by,

\begin{equation}
\label{eq:4}
\text{Pr} \left(\vec{X}(t+1) = \vec{x}+e_{ij}|\vec{X}(t)=\vec{x} \right) =\nu_{ij}(\vec{x}) \quad \text{for} \quad i\neq j, 
\end{equation}

\begin{equation}
\label{eq:5}
\text{Pr} \left(\vec{X}(t+1) = \vec{x}|\vec{X}(t)=\vec{x} \right) = 1- \sum_{i\neq j} \nu_{ij}(\vec{x}).
\end{equation}

\subsection{The Master Equation}

Using the stochastic dynamics of the system, we can calculate the master equation for the probability distribution $P( \vec{x},t)=P(\vec{X}(t)=\vec{x})$. We have,

\begin{equation}
    \label{eq:6}
    P(\vec{x},t+1) = P(\vec{x},t) + \sum_{i \neq j}\left[\nu_{ij}(\vec{x}-e_{ij})\,P( \vec{x}-e_{ij}, t)-\nu_{ij}(\vec{x})\,P(\vec{x},t)\right].
\end{equation}

\section{The Continuum Model}
\label{sec3}
To obtain a continuum model, we extend the previous model as follows: Let $\vec{x}=(x_1,x_2, \dots, x_n)$ be the state of the system at time $t$ and $l>0$ be the step size. At time step $t+1$, two agents $i$ and $j$ are selected at random, and agent $i$ consumes $l$ amount of agent $j$ with some given probability $\kappa_{ij}(\vec{x})$ which is zero when either $x_i=0$ or $x_j=0$. 

Assuming that $x_i^0$ is the initial value of $x_i$, the state of the Markov chain is the set 

\begin{equation}
    S=\{(x_1,\dots,x_n):\, x_i \in x_i^0 + l\,\mathbb{Z},\, x_i \geq 0 \}.
\end{equation}

Denote $\nu_{ij}$ as in equation \eqref{eq:3}, and now the stochastic dynamics of the new discrete-state discrete-time Markov chain are given by

\begin{equation}
\label{eq:7}
 \text{Pr} \left(\vec{X}(t+1) = \vec{x}+l\,e_{ij}|\vec{X}(t)=\vec{x} \right) =\nu_{ij}(\vec{x}) \quad \text{for} \quad i\neq j, 
\end{equation}

\begin{equation}
\label{eq:8}
\text{Pr} \left(\vec{X}(t+1) = \vec{x}|\vec{X}(t)=\vec{x} \right) = 1- \sum_{i\neq j} \nu_{ij}(\vec{x}).
\end{equation}

And the corresponding master equation is given by

\begin{equation}
    \label{eq:9}
    P(\vec{x},t+1) = P(\vec{x},t) + \sum_{i\neq j}\left[\nu_{ij}(\vec{x}-l\,e_{ij})\,P(\vec{x}-l\,e_{ij}, t)-\nu_{ij}(\vec{x})\,P(\vec{x},t)\right].
\end{equation}

Assuming that $l$ is small enough, we use the second-order Taylor expansion for $\nu P$:

\begin{equation}
\label{eq:10}
\begin{split}
    &\nu_{ij}(\vec{x}-l\,e_{ij})\,P(\vec{x}-l\,e_{ij},t)\\
    &= \nu_{ij} P(\vec{x},t) -l \,\frac{\partial \nu_{ij} P}{\partial x_i}(\vec{x},t) + l\, \frac{\partial \nu_{ij} P}{\partial x_j}(\vec{x},t) + l^2\,\left( \frac 12\,\frac{\partial^2 \nu_{ij} P}{\partial x_i^2}(\vec{x},t) + \frac 12\,\frac{\partial^2 \nu_{ij} P}{\partial x_j^2}(\vec{x},t) - \frac{\partial^2 \nu_{ij} P}{\partial x_i \, \partial x_j}(\vec{x},t)\right).
\end{split}
\end{equation}
Now, assume that jumps occur in a small time interval $\dt$. Hence, from equations \eqref{eq:9} and \eqref{eq:10}  we obtain the following Fokker-Planck equation for the transient probability distribution $P(\vec{x},t)=P(\vec{X}(t)=\vec{x})$:

\begin{equation}
    \label{eq:11}
     \frac{\partial }{\partial t} P(\vec{x},t) = -\frac l\dt\,\sum\limits_{i=1}^{n} \frac{\partial}{\partial x_i}(a_i(\vec{x})\,P(\vec{x},t))+ \frac{l^2}{2\,\dt}\,\sum\limits_{i=1}^{n} \frac{\partial^2}{\partial x_i^2}(b_i(\vec{x})\,P(\vec{x},t)) -\frac{l^2}{\dt}\, \mathop{\sum\sum}\limits_{i<j} \frac{\partial^2}{\partial x_i \partial x_j}(c_{ij}(\vec{x})\,P(\vec{x},t)),
\end{equation}     

where 

\begin{equation}
\label{eq:12}
   a_i(\vec{x}) = \mathop{\sum\limits_{j=1}^{n}}_{j \neq i} \left(\nu_{ij}(\vec{x})-\nu_{ji}(\vec{x})\right), \qquad b_i(\vec{x}) = \mathop{\sum\limits_{j=1}^{n}}_{j \neq i} \left(\nu_{ij}(\vec{x})+\nu_{ji}(\vec{x})\right), \qquad c_{ij}(\vec{x}) =\nu_{ij}(\vec{x})+\nu_{ji}(\vec{x}). 
\end{equation}

In the next sections, we study two-agent and multi-agent systems with several models for the rate function $\nu_{ij}$. 
\subsection{A Random Walk On a Straight Line}
We consider a model with $n=2$. In this case, the model is a random walk on the straight line $x+y=N$  with two end-points $(0, N)$ and $(N,0)$ where $N=X_1(t)+X_2(t)$ is constant over time.
\subsubsection{A Symmetric Random Walk}
 We assume that the process is symmetric, i.e., for some positive constant $c$,


\begin{equation}
\label{eq:13}
\nu_{1,2}(x_1,x_2) = \nu_{2,1}(x_1,x_2)=
\begin{cases}
 c, & \text{if $x_1\,x_2>0$}\\
0, & \text{if $x_1\,x_2=0$}.
\end{cases}
\end{equation}

 Now let $\dt=l^2$ and take $\dt \to 0$. From equation \eqref{eq:11} we have the following diffusion equation for $f(x_1,t|x_1^0,t_0)=P(t,x_1,N-x_1| x^0, t_0)$:

\begin{equation}
\label{eq:14}
\begin{split}
     &\frac{\partial }{\partial t} f(x_1,t| x_1^0, t_0) =  4\,c\, \frac{\partial^2}{\partial x_1^2}f(x_1,t| x_1^0, t_0).
\end{split}
\end{equation}


The Guassion solution to \eqref{eq:14} with the initial condition $f(x_1,t_0|x_1^0,t_0)=\delta(x_1-x_1^0)$ is given by

\begin{equation}
\label{eq:15}
Q(x_1,t | x_1^0, t_0)=
C\,\frac{1}{\sqrt{t-t_0}}\, e^{-\frac{(x_1-x_1^0)^2}{16\,c\,(t-t_0)}}.
\end{equation}

Now we consider two sticky bounderies at $\alpha=(0,N)$ and $\beta=(N, 0)$. We can write the above diffusion equation with the following boundary conditions,

\begin{equation}
    \label{eq:16}
    \begin{split}
        &\frac{\partial }{\partial t} f(x_1,t| x^0, t_0) =  4\,c\, \frac{\partial^2}{\partial x_1^2}f(x_1,t| x^0, t_0),\\
     & f(x_1,t_0|x^0, t_0) = \delta(x_1-x_1^0).\\
     &f(x_1,t|0,t_0) = \delta(x_1),\\
    &f(x_1,t|N,t_0) = \delta(x_1-N).
    \end{split}
\end{equation}

The last two conditions in diffusion equation \eqref{eq:16} are given due to the fact that once the process reaches the states $\alpha$ or $\beta$ it remains there forever. We call $\alpha$ and $\beta$ the sticky boundaries. 
First, we consider a process with natural boundaries, and the solution is given by the Gaussian formula in \eqref{eq:15} which satisfies the initial condition. To modify the solution for the boundary conditions given in \eqref{eq:16}, we apply the image method and we have,

\begin{equation}
\label{eq:17}
     f(x_1,t|x_1^0,t_0) =  Q(x_1,t|x_1^0,t_0) - u(0,N,x_1^0)\,\left(Q(x_1,t|0,t_0)-\delta(x_1) \right) - v(0,N,x_1^0)\,\left(Q(x_1,t|N,t_0)-\delta(x_1-N) \right).
\end{equation}

Here $u$ and $v$ are constants with

\begin{equation}
\label{eq:18}
\begin{split}
    &u(0,N,0)=1, \qquad u(0,N,N)=0,\\ &v(0,N,0)=0, \qquad u(0,N,N)=1.
\end{split}
\end{equation}

The constants $u$ and $v$ are actually the probability of reaching states $\alpha$ and $\beta$ respectively. By symmetry, if the process starts at $x^0=(x_1^0, N-x_1^0)$, then  we can derive the following equations:

\begin{equation}
\label{eq:19}
\begin{split}
 u(0, N, x_1^0) = \left|\frac{N-x_1^0}{N}\right|, \qquad
v(0, N, x_1^0) = \left|\frac{x_1^0}{N}\right|.
\end{split}
\end{equation}

There is a simple interpretation for equation $\eqref{eq:17}$. If we rearrange the terms, then we have,

\begin{equation}
\label{eq:20}
\begin{split}
    f(x_1,t|x_1^0,t_0) &=\;   Q(x_1,t|x_1^0,t_0) - u(0,N,x_1^0)\,Q(x_1,t|0,t_0)- v(0,N,x_1^0)\,Q(x_1,t|N,t_0)\\
    &+ u(0,N,x_1^0)\,\delta(x_1) + v(0,N,x_1^0)\,\delta(x_1-N).
\end{split}
\end{equation}

The first term on the right-hand side of equation \eqref{eq:20} describes a diffusion process with a natural boundary that travels on the whole line and is unaware of the boundaries. The loss of particles is described in the second and third terms where particles are removed by the absorbing walls at $x=\alpha$ and $x=\beta$. But, since the system is conservative, the loss of particles is corrected by the last two terms (delta functions) which describe the particles that are glued to the walls and remain there forever.\\

\subsection{Random Walk on a Plane}

In this section, we consider a 3-agent model which describes a random walk on the plane $x_1+x_2+x_3=N$. From the definition of the model, we have three elastic boundaries $(0,x_2,N-x_2)$, $(x_1,0,N-x_1)$, $(x_1,N-x_1,0)$. These boundaries are straight lines and once the particle reaches these boundaries it remains in those boundaries and can only move there. If the particle reaches the points $(N,0,0)$, $(0,N,0)$ or $(0,0,N)$ it remains there foreover. 

With the constant transition rate
\begin{equation}
\label{eq:21}
\nu_{i,j}(x_i,x_j)= 
\begin{cases}
 c, & x_i\,x_j>0,\\
0, & x_i\,x_j=0,
\end{cases}
\end{equation}

we have the following diffusion equation from \eqref{eq:11},

\begin{equation}
\label{eq:22}
\begin{split}
     &\frac{\partial }{\partial t} f(x,t| x^0, t_0) =  4\,c\, \frac{\partial^2}{\partial x_1^2}f(x,t| x^0, t_0) + 4\,c\, \frac{\partial^2}{\partial x_2^2}f(x,t| x^0, t_0) + 4\,c\, \frac{\partial^2}{\partial x_1\,\partial x_2}f(x,t| x^0, t_0),\\
     & f(x,t_0|x^0,t_0) = \delta(x-x^0).
\end{split}
\end{equation}

The solution to the above Fokker-Planck equation with the initial condition $Q(x,t_0|x^0,t_0)=\delta(x-x^0)$ is given by

\begin{equation}
\label{eq:23}
    Q(x,t|x^0,t_0) = 
    C \, \frac{1}{t - t_0} \, \exp\left( - \frac{(x_1 - x_1^0)^2 + (x_2 - x_2^0)^2 + 4(x_1 - x_1^0)(x_2 - x_2^0)}{32 \, c \, (t - t_0)} \right).
\end{equation}

Considering the elastic boundaries described above, the special solution with the boundary conditions is 


\begin{equation}
\label{eq:24}
   \begin{split}
       f(x,t|x^0,t_0) &=  Q(x,t|x^0,t_0) - w_1(x^0)\,\left(Q(x,t|0,x_2^0,t_0) - g_1(x_2)\right) \\
       & - w_2(x^0)\,\left(Q(x,t|x_1^0,0,t_0) - g_2(x_1)\right) \\
       & - w_3(x^0)\,\left(Q(x,t|x_1^0,N - x_1^0,t_0) - g_3(x_1)\right) \\
       & - w_{12}(x^0)\,\left(Q(x,t|0,0,t_0) - \delta(x_1,x_2)\right) \\
       &- w_{13}(x^0)\,\left(Q(x,t|0,N,t_0) - \delta(x_1,x_2 - N)\right) \\
       & - w_{23}(x^0)\,\left(Q(x,t|N,0,t_0) - \delta(x_1 - N,x_2)\right).
   \end{split}
\end{equation}

where



\begin{equation}
\label{eq:25}
    \begin{split}
        g_1(x_2) &= Q^*(x_2,t|x_2^0,t_0) - u(0,0,N,x_2^0)\,\left(Q^*(x_2,t|0,t_0) - \delta(x_2) \right) \\
        &- v(0,0,N,x_2^0)\,\left(Q^*(x_2,t|N,t_0) - \delta(x_2 - N) \right), \\
        g_2(x_1) &= Q^*(x_1,t|x_1^0,t_0) - u(0,0,N,x_1^0)\,\left(Q^*(x_1,t|0,t_0) - \delta(x_1) \right) \\
        & - v(0,0,N,x_1^0)\,\left(Q(x_1,t|N,t_0) - \delta(x_1 - N) \right), \\
        g_3(x_1) &= Q^*(x_1,t|x_1^0,t_0) - u(0,N,0,x_1^0)\,\left(Q^*(x_1,t|0,t_0) - \delta(x_1) \right) \\
        & - v(0,N,0,x_1^0)\,\left(Q^*(x_1,t|N,t_0) - \delta(x_1 - N) \right).
    \end{split}
\end{equation}

and for $i,j=1,2,3$,

\begin{equation}
    \begin{split}
        \label{eq:26}
        &w_i(x)=
        \begin{cases}
         1 & x_i=0, x_j\neq 0,\\
         0 & x_i \neq 0, x_j=0\\
         0 & x_i=x_j=0
        \end{cases}\\
        &w_{ij}(x)=
        \begin{cases}
         1 & x_i=0 ,\, x_j=0,\\
         0 & x_i=0, x_j \neq 0
        \end{cases}\\
    \end{split}
\end{equation}

\begin{center}
 \includegraphics[scale=0.5]{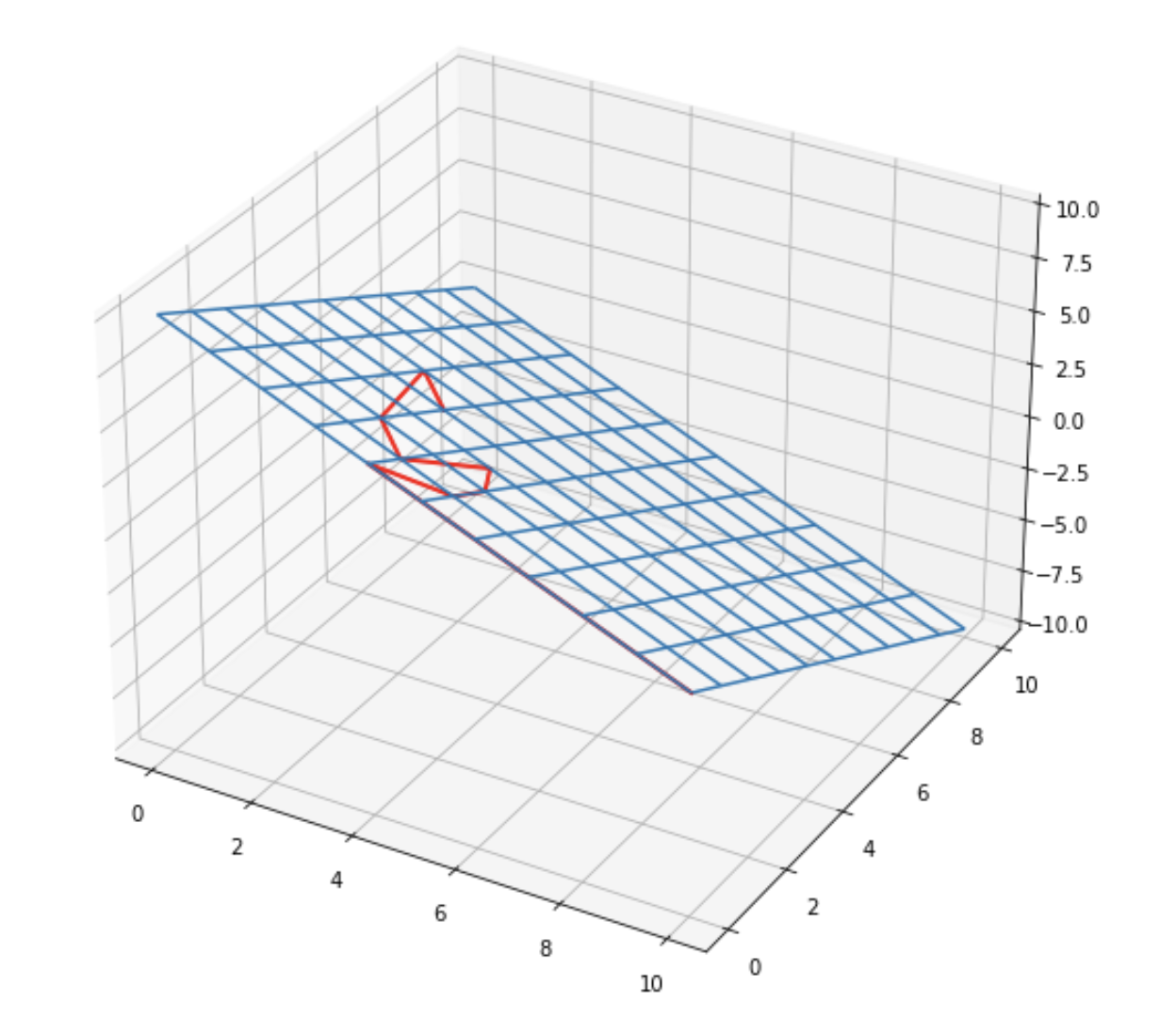}   
 \captionof{figure}{A random walk on $X+Y+Z=10$ with elastic boundaries and sticky corners.}
\end{center}

\section{Conclusions}
In this project, we used continuous-time Markov chains to analyze the behavior of a multi-agent system in a competitive environment. To examine the behavior of agents who exhibit competitive behavior, we introduced the fundamental ideas of Markov chains and created a stochastic agent-based model. Our model has a wide range of applications for modeling stochastic systems where entities struggle for resources or influence, such as in finance, social science, and politics. For specific situations where the rate of consumption is linear, we obtained the Fokker-Planck equations for the transient probability distribution of the system's density and solved them. Our research deepens our understanding of how competitors behave in multi-agent systems and can be applied to the development of winning strategies in competitive settings.


\pagebreak

\bibliographystyle{siam}
\bibliography{refs}

\end{document}